\documentclass[%
 reprint,
 amsmath,amssymb,
 aps,
]{revtex4-1}

\usepackage{graphicx}
\usepackage{dcolumn}
\usepackage{bm}
\usepackage{subcaption}
\usepackage{floatrow}
\usepackage{color}
\newfloatcommand{capbtabbox}{table}[][\FBwidth]
\usepackage[T1]{fontenc}
\usepackage[font=small,labelfont=bf,tableposition=top]{caption}

\DeclareCaptionLabelFormat{andtable}{#1~#2  \&  \tablename~\thetable}

\def\Journal#1#2#3#4{{#1} {\bf #2}, #3 (#4)}

\def\NCI{\em Nuovo Cimento C}

\def\NIMA{{\em Nucl. Instrum. Methods} A}

\def\PLB{{\em Phys. Lett.} B}

\def\JPG{\em Journal of Physics G}
\def\APP{\em Astroparticle Physics} 
\def\APJ{\em The Astrophys. J.}
\def\APJL{\em The Astrophys. J. Lett.}
\def\AAA{\em Astro. and Astrophys.}
\def\AAS{\em Astro. and Astrophys. Suppl.}
\def\CPC{\em Chin. Phys. C}
\def\MNRAS{\em Month Not. Royal Ast. Soc.}
\def\PRL{\em Phys. Rev. Lett.} 
 
\def\NAT{\em Nature} 
\def\SCI{\em Science}

\def\PRD{{\em Phys. Rev.} D} 
\def\PRC{{\em Phys. Rev.} C}

\def\APJ{\em ApJ}

\def\PTEP{\em Prog. Th. Exp. Phys.}

\begin{document}


\title{Prediction and detection potential of fusion neutrinos from nearby stars}

\author{K. Zuber}
\email{zuber@physik.tu-dresden.de}
 \affiliation{Institute for Nuclear and Particle Physics, Technische Universit\"at 
Dresden,  01069 Dresden, Germany}
\author{S. Arceo Diaz}%
 \email{santiagoarceodiaz@gmail.com}
\affiliation{Instituto Tecnol\'ogico de Colima, Colima, M\'exico}%




\date{\today}

\begin{abstract}
The shapes of neutrino spectra and fluxes from representative stars in the solar neighborhood up to 10 pc are calculated.
The individual contribution of the most important, specific stellar objects (the $\alpha$-Centauri system, Sirius A, Procyon, Vega, Fomalhaut and Altair) are determined and investigated by detailed stellar modeling. The possibility for a potential detection is discussed.
\end{abstract}

\pacs{Valid PACS appear here}
\maketitle


\section{Introduction}
\label{sec:intro}
Neutrino astronomy has enormously grown over the last two decades 
and is now a well established field in particle astrophysics. The detection of 
solar neutrinos in real-time in form of direct measurements of the pp
\cite{pp}, pep \cite{pep}, $^7$Be \cite{be7} and $^8$B-neutrinos 
\cite{sno,superk,b8borex} as well as 
radio-chemical observations using $^{37}$Cl \cite{cle98}
and $^{71}$Ga \cite{ham99,alt05,abd09} as a target made the solar energy 
production an experimental topic. Furthermore, the long standing problem of 
missing solar neutrinos has been solved and lead to
neutrino flavor conversion happening within the Sun. Observations of neutrinos from SN 1987A \cite{sn87imb,sn87kam,sn87baksan} confirmed the basic picture of core collapse 
supernova. Recently, the neutrino telescope
ICECUBE provided detection of very high energetic neutrinos from extra-galactic 
sources \cite{aar13}. Also, experimental bounds on other interesting quantities 
like the flux of a diffuse supernova neutrino background have reached 
sensitivities as low as
2-3 events $\mathrm{cm^{-2}\cdot s^{-1}}$ \cite{bay12}.\\ The next generation of underground experiments, Hyper-Kamiokande, a 1 Mt water Cherenkov detector \cite{hyperk}, JUNO and RENO-50, 20 kt liquid scintillation detectors \cite{juno,reno-50} and a potential large scale scintillator experiment deep underground in Jinping\cite{bea16} might allow to detect additional weak neutrino sources besides the Sun.\\ In this paper the opportunity is explored whether neutrinos produced by stars from the solar neighborhood might be detectable in the near future. In the same way as the study of planets has been extended to exoplanets and helioseismology towards asteroseismology, this
would be another new, unique extension of solar system studies out into the 
Milky Way. 
A very general approach predicting galactic neutrinos using global properties 
of 
the Milky Way itself can be found in \cite{bro98}.

\section{Nearby stars}
\label{sec:nearby}
A search for extra solar sources of fusion neutrinos should be dominated from nearby stars due to the flux dependence on distance. For this analysis, the solar neighborhood has been restricted to 10 pc. Most the stars within this distance are red dwarfs, approximately 215 out of 321 (figure 1) according to the observational data in the Gliese catalogue \cite{GLI}, showing a large spread in their metallicity and mass. The contribution of most of these stars to any neutrino flux can be assumed to be negligible (see below).  

\begin{figure}
\begin{center}
\label{fig:cmdiagram}
\includegraphics[scale=0.20]{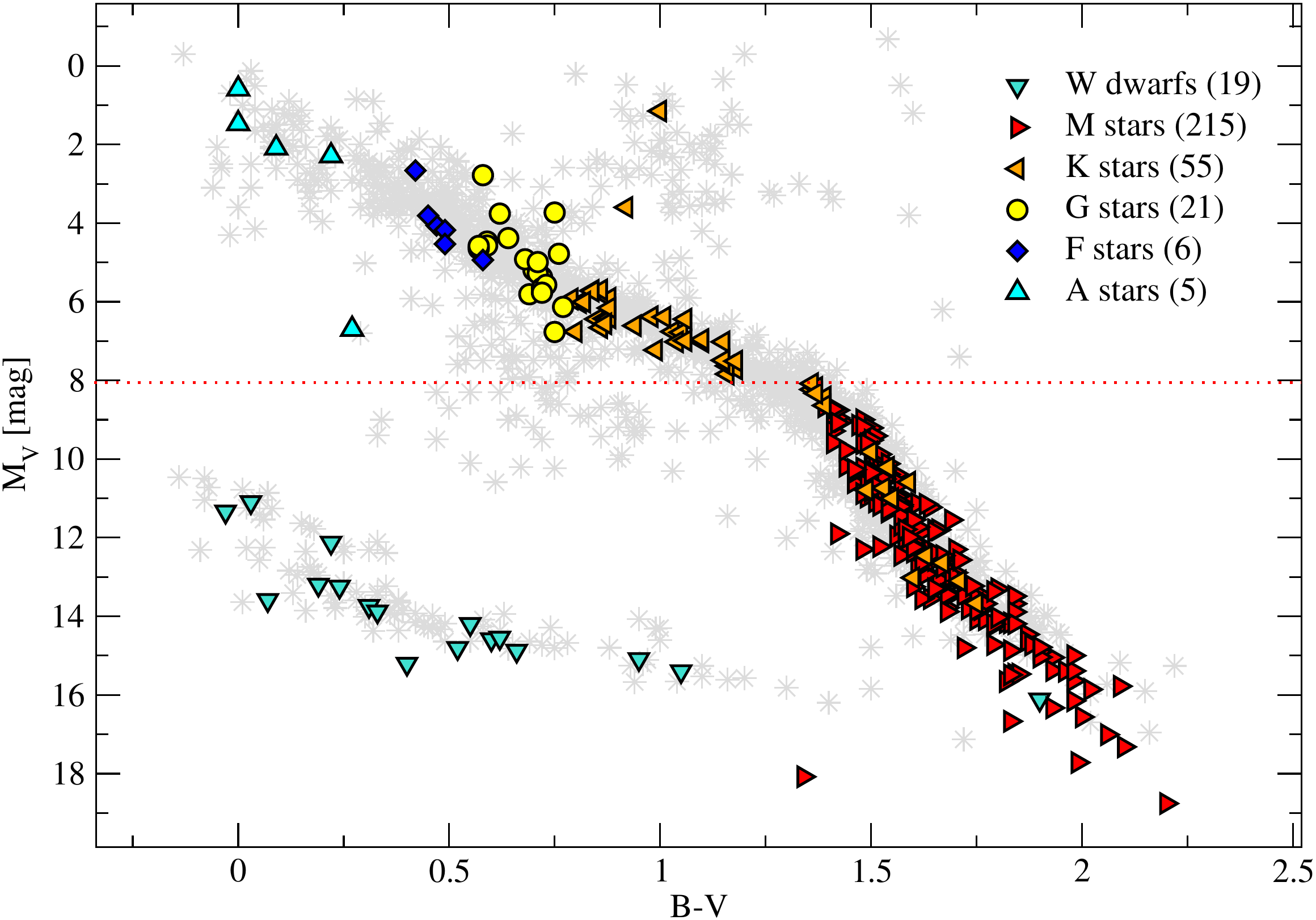}
\protect\caption[]{\footnotesize{Color-magnitude diagram (CMD) of the stars in the Gliese catalogue. Stars within 10 pc are separated from the background (grey marks). The horizontal line corresponds to the theoretical threshold below which neutrino 
luminosity should be less than 10\% of the solar value.}}
\end{center}
\label{figure1}
\end{figure}

The Eggleton code for stellar evolution \cite{Egg} was used to simulate each spectral class within the solar neighborhood, most of its characteristics according to the version described in \cite{pols95} and modified in \cite{Arceo2015} to include more recent tables for plasmon decay \cite{Haft94}, one of the thermal reactions producing neutrinos \cite{Itoh92}. A simulation of the Sun, taking Z=0.02, resulted in a 
core temperature of $\mathrm{T_{c}}$ = 15.76 (in units of $10^{6}$ K) and density $\rho_{c}=1.60$ (in units of $\mathrm{10^{2}g\cdot cm^{-3}}$) at the point in which the stellar track achieves the bolometric luminosity of the Sun, a difference smaller than 1\% with respect to standard solar models, for which $\mathrm{T_{c}=15.70}$  and $\mathrm{\rho_{c}=1.62}$ \cite{BP2004,bah05}. For all the models shown throughout this work, the effective temperature, radius and luminosity differ from their observational counterparts by around 5\%, the effective temperature usually showing the largest deviation.

As a condition for nearby stars to be considered relevant for the extra-solar 
neutrino flux at Earth, it was required that the minimum neutrino luminosity of their models should be at least
$10\%$ of the solar neutrino output. According to our numerical simulations, and the empirical bolometric corrections by \cite{hab81}, any main-sequence star with a visual magnitude $\mathrm{M_{V}<18}$ can be considered as a relevant neutrino source, corresponding to a stellar mass around $\mathrm{0.6M_{\odot}}$ (the exact value varies slightly with metallicity: $\mathrm{M_{i}=0.63M_{\odot}}$ for 
$\mathrm{Z=0.001}$ and $\mathrm{M_{i}=0.59M_{\odot}}$ for $\mathrm{Z=0.02}$). This threshold is marked by the dotted line in the CMD shown in figure 1, ruling out all the M dwarfs and most main-sequence K stars.

\section{The neutrino spectra from the strongest sources within 10 pc}

Seven stars were selected, based on their intrinsic brightness and proximity as the dominant sources for extra solar neutrinos (table I).  On each case, the observational estimations on mass and metallicity were used as an input for the stellar tracks, while the bolometric luminosity, effective temperature, radius and age, were set as references to compare with. Once the observational data were matched, the internal physical conditions on the stellar models were used to calculate the neutrino fluxes and spectra shown below.

\begin{table}
\label{tab:tab:4}
\scriptsize
\begin{tabular}{|l c c c |}
	\hline
\textbf{A star}   & distance [pc] & \textbf{F star}   & distance [pc]   \\
\hline
Sirius    & 2.64 &  Procyon  & 3.50 \\
Altair      & 5.15  & Tabit  & 8.03  \\
Alpha Fomalhaut   & 7.69  & Chi Draconis & 8.06  \\
Vega  & 7.75 & Gamma Leporis  & 8.97 \\
\hline 
\textbf{G star}   & distance [pc] & \textbf{K star}   & distance [pc]   \\
\hline
Alpha Centauri A & 1.35 & Alpha Centauri b & 1.35\\
Tau Ceti & 3.65  & Epsilon Eridani B & 3.22\\
Eta Cassiopeiae & 5.95 & 61 Cyg & 3.48\\
82 Eridani & 6.06 & Struve 2398 & 3.51 \\
Delta Pavonis & 6.11 & Gliese 725A & 3.51\\
Xi Bootis & 6.70 & G Epsilon Indi & 3.62\\
\hline
\end{tabular}
\protect\caption[]{\footnotesize{Closest neutrino sources within 10 pc arranged according to spectral classes, A (upper left), F (upper right), 
G (lower left) and K (lower right). The brightest and closest sources, therefore providing the largest contribution to the flux, are shown in tables II and III.}}
\end{table}

\begin{table*}
\label{tab:tab:4}
\scriptsize
\begin{tabular}{|l c c c c c c c|}
	\hline
 \textbf{Star} & $\mathrm{\alpha-Cen_{A}}$ & $\mathrm{\alpha-Cen_{B}}$ & Procyon & Sirius & Fomalhault & Altair & Vega \\
\hline 
 & &  & Observations &  & & &  \\
 \hline
Age [$10^{9}$yrs]              & $4.850\pm0.5$   & $4.850\pm0.5$  & $1.87\pm0.13$    & $0.225\pm0.025$ & $0.44\pm0.04$  & 1.26  &$0.455\pm 0.0013$\\
$\mathrm{M [M_{\odot}]}$       & $1.100\pm0.006$ & $0.907\pm0.006$& $1.499\pm0.031$  & $2.02\pm0.03$ & $1.92\pm0.02$  & $1.79\pm0.018$  & $2.135\pm0.074$  \\
$\mathrm{T_{eff} [K]}$         & $5790\pm30$     & $5260\pm50$    & $6530\pm50$      & $9940\pm200$        & $8590\pm73$  & $6900-8500$  & $8152-10060$ \\
$\mathrm{L_{bol} [L_{\odot}]}$ & $1.519\pm0.018$ &$0.5002\pm0.016$& $6.92\pm0.05$    & $25.4\pm1.3$  & $16.63\pm0.48$  & $10.60\pm1.02$  & $40.12\pm 0.45$ \\
$\mathrm{R_{*} [R_{\odot}]}$   & $1.23\pm0.003$  & $0.863\pm0.005$& $2.098\pm0.025$  & $1.711\pm0.014$ & $1.842\pm0.019$  & $1.63-2.03$  & $2.81-2.93$  \\
\hline
 & &  & Stellar models &  & & &  \\
 \hline
Age [$10^{9}$yrs]               & 5.24  & 5.26  & 2.06  & 0.26 & 0.45  & 1.36  & 0.66\\
$\mathrm{M [M_{\odot}]}$        & 1.10  & 0.90  & 1.49  & 2.10 & 1.95  & 1.79  & 2.20  \\
$\mathrm{T_{eff} [K]}$    & 5840  & 5180  & 6500  & 9780 & 8300  & 7860  & 3930  \\
$\mathrm{L_{bol} [L_{\odot}]}$  & 1.52  & 0.46  & 6.92  &24.61 &14.73  &10.57  &39.87 \\
$\mathrm{R_{*} [R_{\odot}]}$    & 1.23  & 0.85  & 2.08  & 1.73 & 1.86  & 1.76  & 2.96  \\
$\mathrm{T_{c} [10^{6} K]}$    & 18.74 & 13.44 & 26.44 & 22.25 & 21.53 &18.23 &28.44 \\
$\mathrm{\rho_{c} [10^{2} g cm^{-3}]}$ &  2.35 &  1.12 &  4.09 &  0.68 & 0.72 & 0.59 & 1.16 \\
$\mathrm{L_{\nu T}[L_{\nu\odot}]}$ &  1.97 & 0.33 & 14.59 & 61.37& 35.40 & 9.54 & 83.35 \\
$\mathrm{L_{\nu pp}[L_{\nu T}]}$     & 0.654 & 0.963 & 0.251 & 0.299& 0.306 & 0.272 & 0.056 \\
$\mathrm{L_{\nu CNO}[L_{\nu T}]}$    & 0.345 & 0.028 & 0.748 & 0.700& 0.693 & 0.727 & 0.943  \\
$\mathrm{L_{\nu th}[L_{\nu T}]}$     & 3.63(-7) & less than (-8)&  1.23(-7)& 2.93(-7) & 1.82(-7) & 1.83(-7)& 2.23(-6) \\
\hline
\end{tabular}
\protect\caption[]{\footnotesize{On each column: stellar parameters and neutrino luminosity for 
our selected models. The numbers inside parenthesis represents powers of 10. The references to the observational calibrations are given in the text.}}
\end{table*}

\subsubsection{$\alpha$-Centauri}

This triple stellar system is the closest to Earth. Each one of its members: $\alpha$-Centauri A (spectral class G2V), 
$\alpha$-Centauri B (spectral class K0V) and Proxima-Centauri (a red dwarf whose mass falls below the threshold) offers the opportunity of comparing their neutrino output to that of the Sun and to analyze how it depends on spectral class, initial mass and chemical composition. 
Table II shows the observable parameters and stellar models for the main components of the $\alpha$-Centauri system, the estimated stellar parameters are in good agreement with the calibrations obtained by interferometry and astroseismology. The stellar models predict that due to the mass difference of the two stars, resulting in different  temperature and density in their cores, the neutrino luminosity varies  from just one third ($\alpha$-Centauri B) to twice the solar value for $\alpha$-Centauri A. 
The reactions from the pp-chain are the main sources of neutrinos (96\% of the total neutrino luminosity in $\alpha$-Cen B and 65\% for A). In addition, according to the simulations the neutrino luminosity due to thermal reactions is less than 1\%.\\
The two upper panels in fig. 2 show the calculated neutrino spectrum for $\alpha$-Centauri A and $\alpha$-Centauri B as if the stars were placed at 1 AU from Earth, which allows a comparison with the solar neutrino spectrum. 
The neutrino flux from the two main stars adds up to 1.5 times the solar production, each emitting $99\%$ and 80\% of their fluxes by the pp-I reaction. The hotter core in $\alpha$-Centauri A should enhance the reactions of the CNO-cycle, making the corresponding CNO-neutrino flux an order of magnitude larger than solar. The flux from the pp-I and p-e-p reactions are maintained at the same level as in the Sun. It appears that, as a consequence on the enhancement of the CNO-cycle, the $\mathrm{^{7}Be}$ and $\mathrm{^{8}B}$ neutrino fluxes should be less reduced (two orders of magnitude lower in this case). The overall neutrino spectrum for $\alpha$-Centauri B is very similar to that of the Sun. However, there is a difference in the fluxes from the $\mathrm{^{8}B}$ and $\mathrm{^{7}Be}$ reactions as on $\alpha$-Centauri B they are several orders of magnitude smaller if treating $\alpha$-Centauri B just as an identical, but dimmer version of the Sun. The reduced neutrino flux coming from the $\mathrm{^{8}B}$ could be a consequence of the high sensitivity to energy generation on temperature ($\mathrm{\propto T^{25}}$), having the highest exponent among the reactions considered in this work \cite{Bah96}.

\subsubsection{Procyon A}

Procyon A (spectral class F5IV) is the brightest member in the $\alpha$-CMi system, along with a white dwarf whose mass is just below the proposed threshold. Most of Procyon's observable parameters, (upper part on table II) have been obtained by astroseismological measurements and allow to classify it as young red giant \cite{Proc1, Proc2}. The model for Procyon is shown in the third column on the lower part of table II. Unlike the other stars considered in this work, nuclear reactions take place within an hydrogen-burning shell surrounding Procyon's core, in which electron degeneracy is already developing (as can be inferred from its high density and temperature). 
The conditions inside the hydrogen burning shell are more favorable for the CNO-cycle, producing 75\% of the total neutrino luminosity (14.6 times the solar value). Due to the increasing degeneracy of the core, plasmon decay substitutes as the dominant source for thermal neutrinos (although the resulting luminosity should be still far smaller than that of nuclear reactions).
The theoretical neutrino spectrum of Procyon A, if the star was located at 1 AU from Earth, is displayed on the left middle panel in figure 2. The integrated flux ($\mathrm{\Phi_{T}=103\times10^{10}\cdot cm^{-2}\cdot s^{-1}}$) is about eighteen times larger than that of the standard solar model. Each reaction in the CNO-cycle is enhanced by three orders of magnitude. Even when neutrino production by the pp-chain is less important than the CNO-cycle, the neutrino flux by the ppI-reaction is almost an order of magnitude larger than that of the standard solar model, with the $^{8}$B and pep-reactions being almost equal. 

\begin{figure*}
\centering
\begin{subfigure}[b]{0.40\linewidth}
\includegraphics[width=1.0\linewidth]{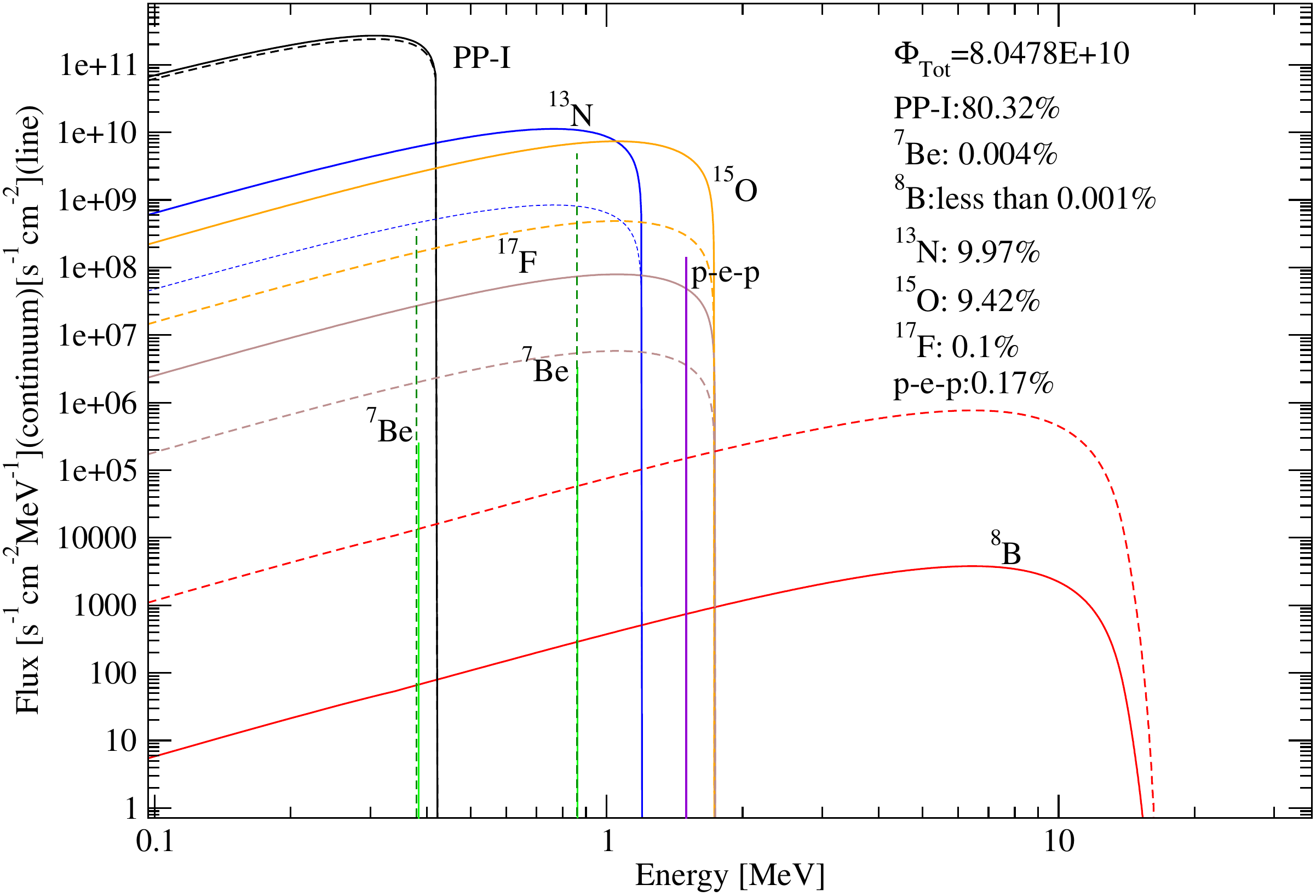}
\caption{\ensuremath{\alpha}-Centauri A\label{sfig:testa}}
\end{subfigure}
\begin{subfigure}[b]{0.40\linewidth}
\includegraphics[width=1.0\linewidth]{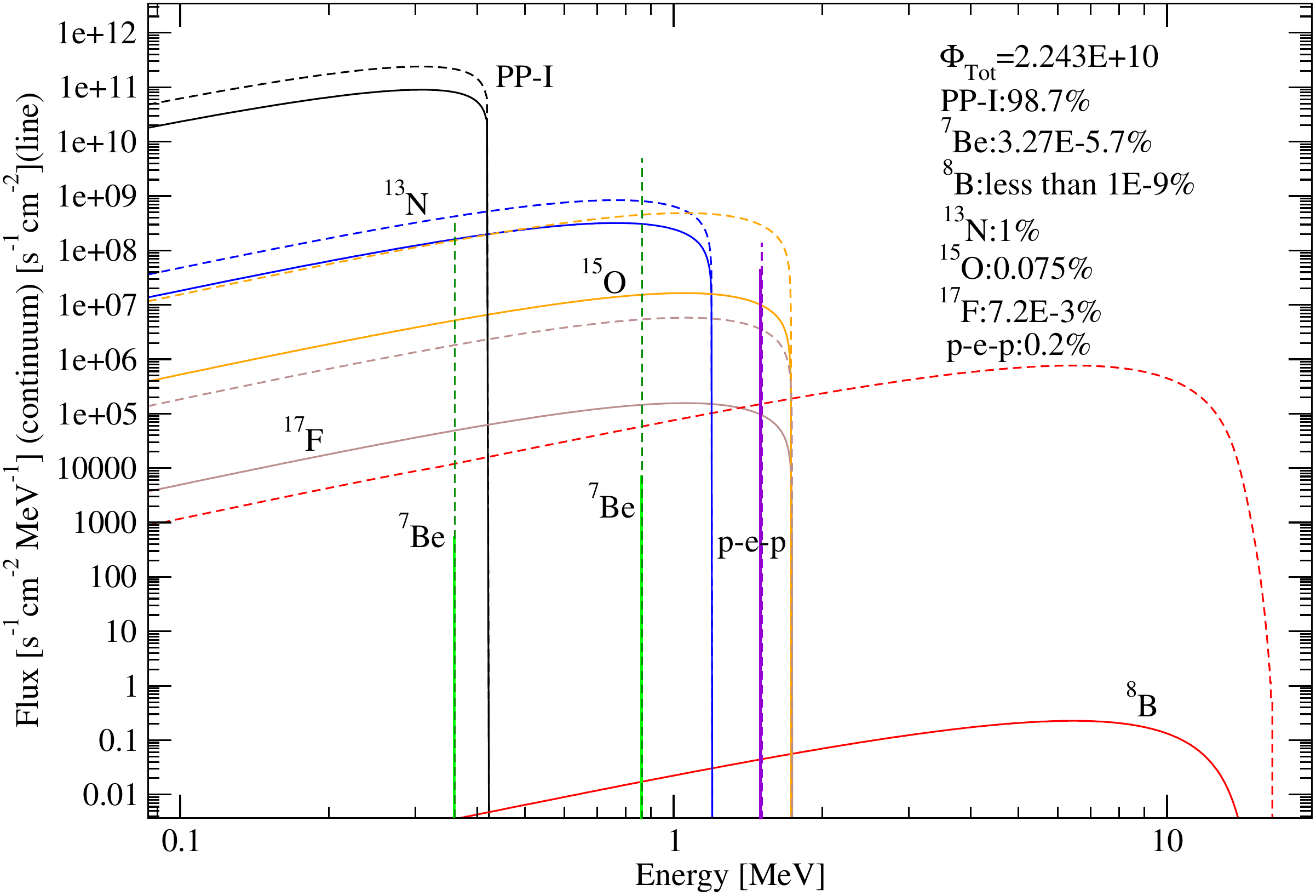}
\caption{\ensuremath{\alpha}-Centauri B\label{sfig:testa}}
\end{subfigure}\\
\begin{subfigure}[b]{0.40\linewidth}
\includegraphics[width=1.0\linewidth]{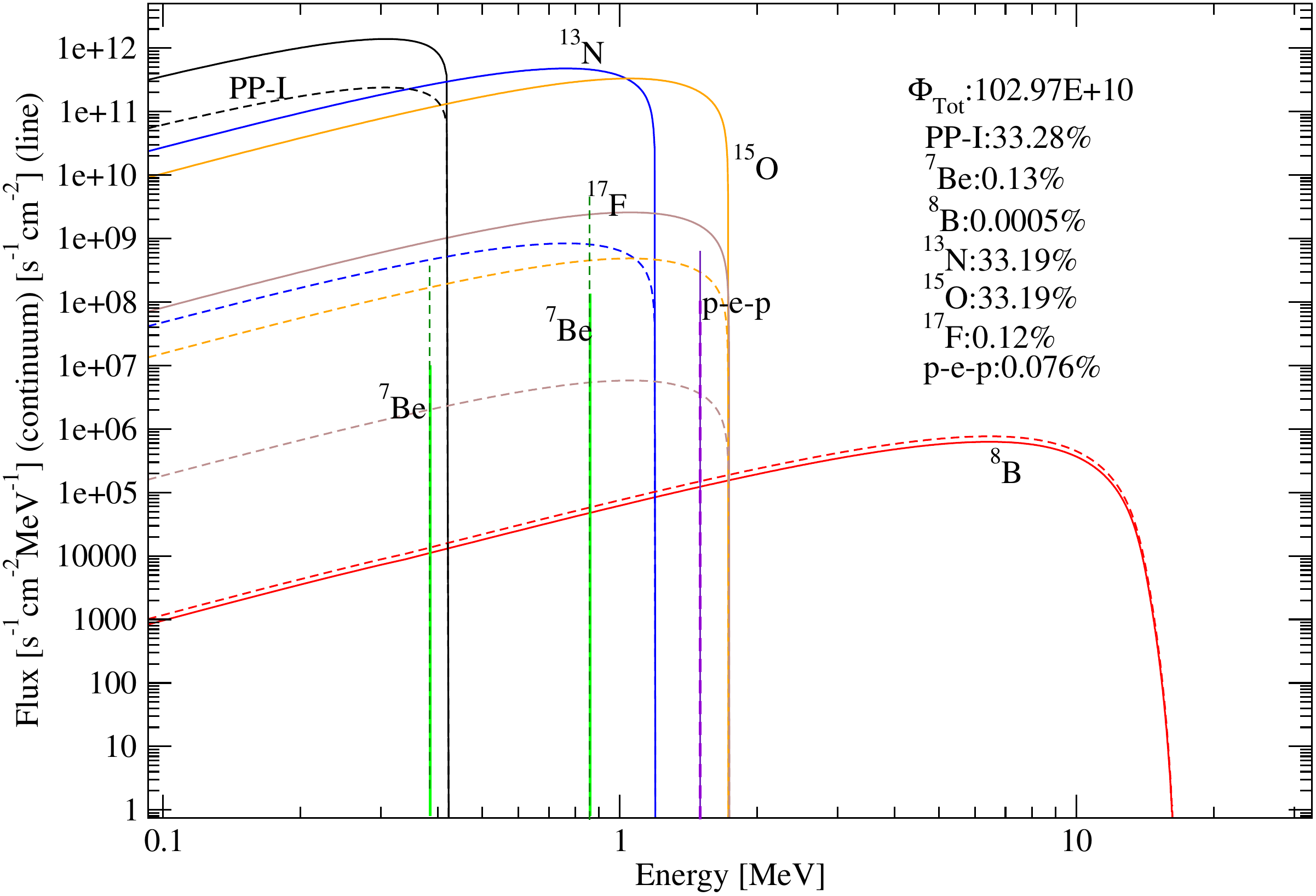}
\caption{Procyon}
\end{subfigure}
\begin{subfigure}[b]{0.40\linewidth}
\includegraphics[width=1.0\linewidth]{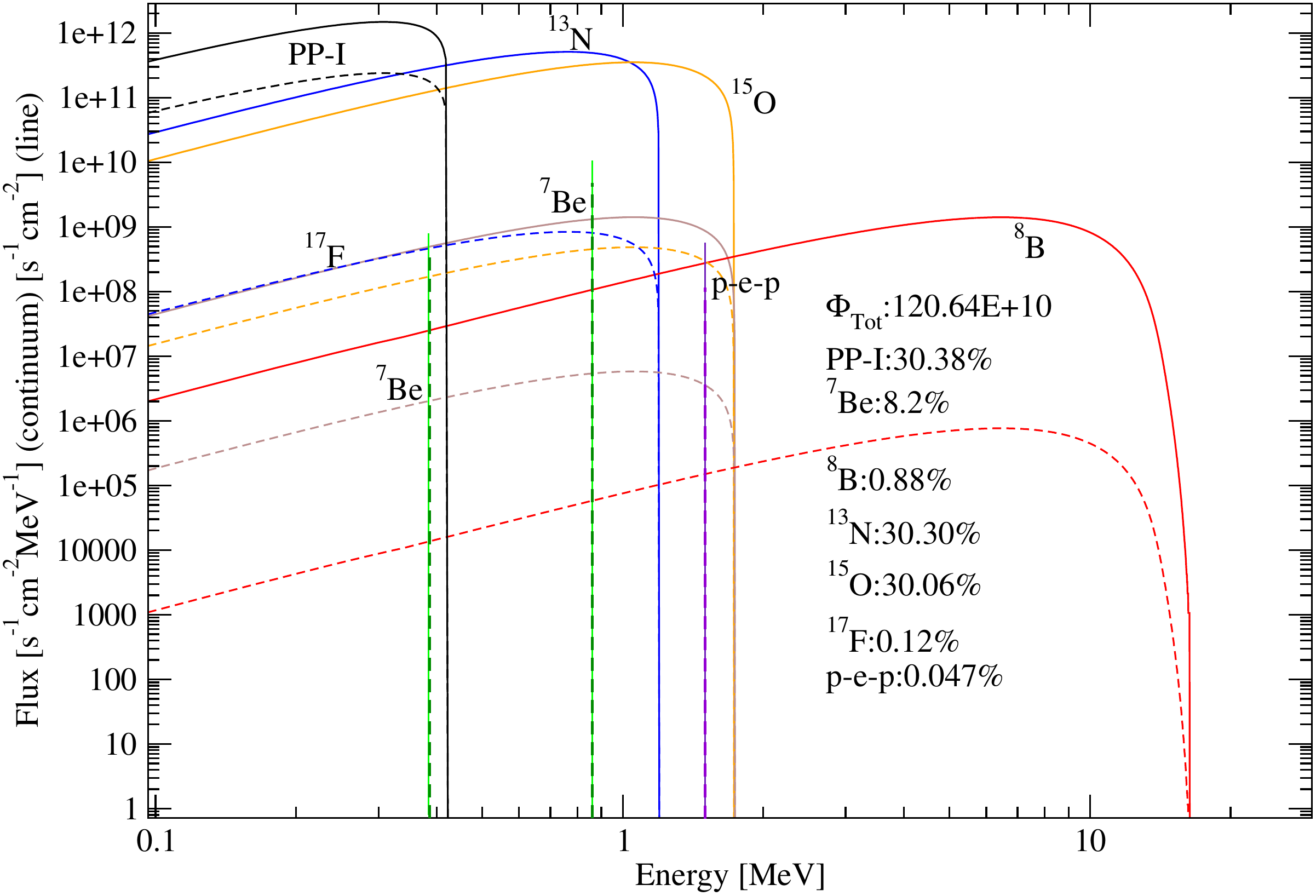}
\caption{Sirius}
\end{subfigure}\\
\begin{subfigure}[b]{0.40\linewidth}
\includegraphics[width=1.0\linewidth]{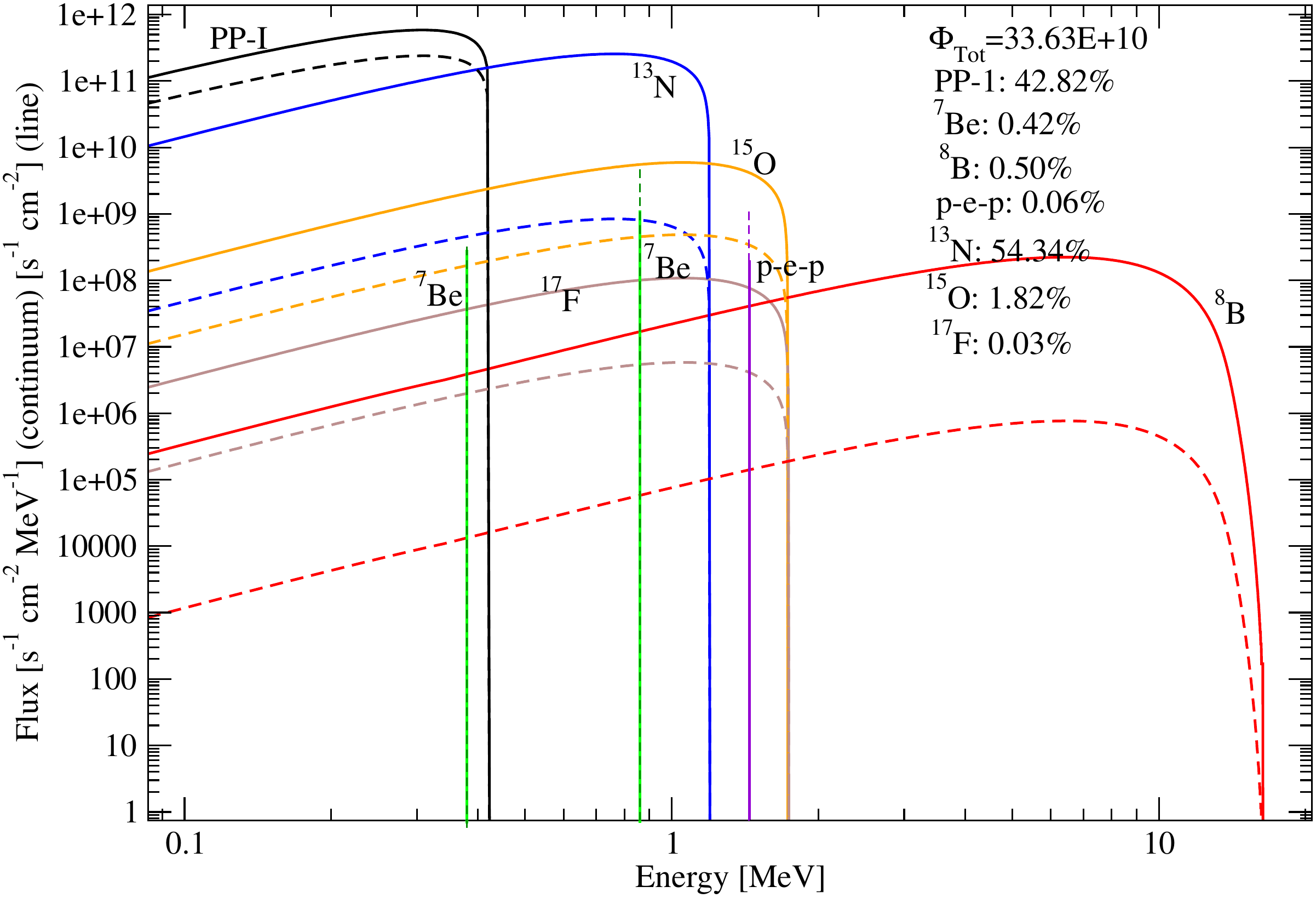}
\caption{Altair}
\end{subfigure}
\begin{subfigure}[b]{0.40\linewidth}
\includegraphics[width=1.0\linewidth]{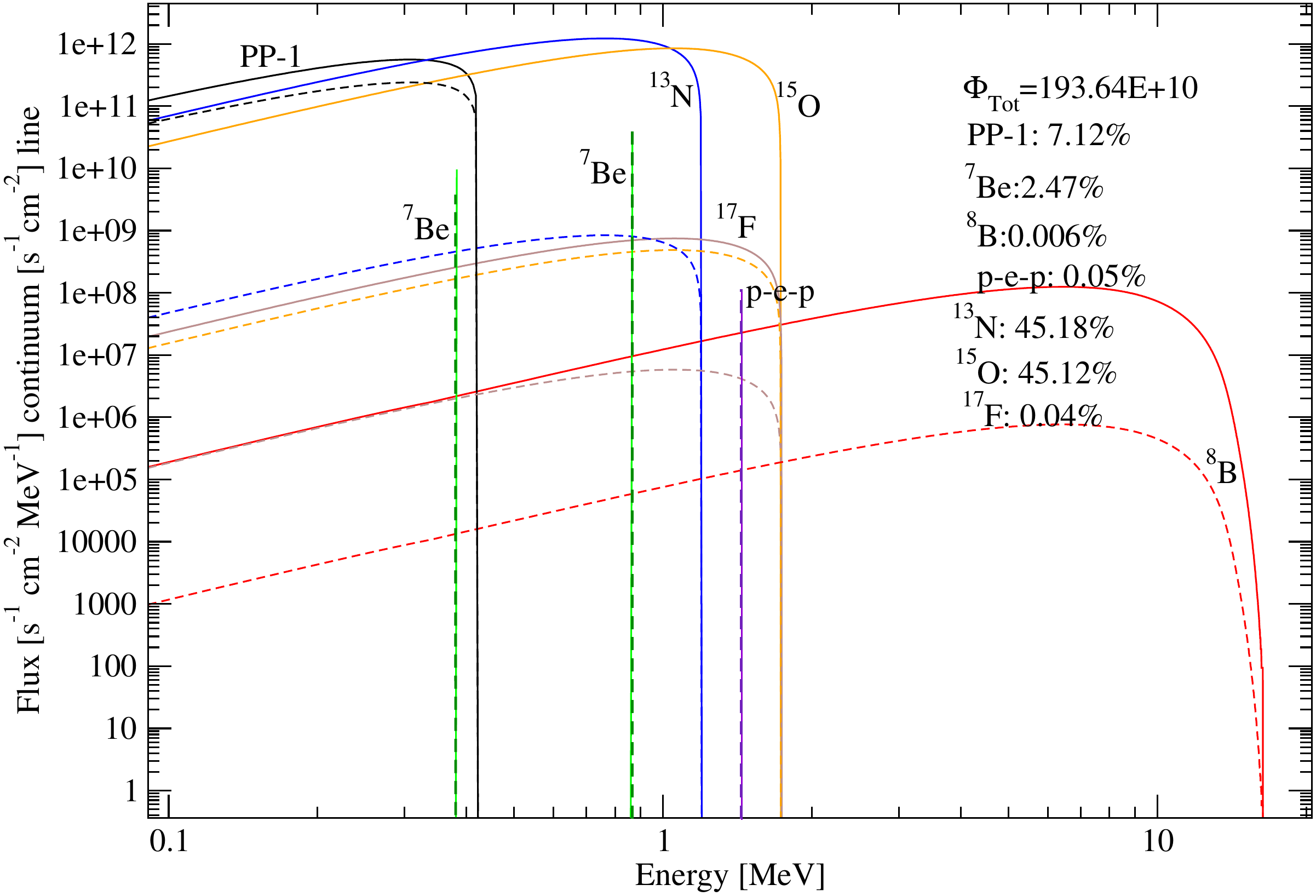}
\caption{Vega}
\end{subfigure}\\
\scriptsize
\begin{tabular}{|l c c c c c c c r|}
	\hline
Flux & BP04  & $\mathrm{\alpha-Centauri_{A}}$ & 
$\mathrm{\alpha-Centauri_{B}}$ & Procyon & Sirius & Altair & Fomalhaut & Vega  \\
$[\mathrm{10^{9}s^{-1}\cdot cm^{-2}}]$ & & & & & & & & \\
\hline 
$\mathrm{\Phi_{T}}$     & 5.96(1)  & 8.05(1)  & 2.24(1)   & 1.03(3) & 1.21(3) &3.36(1)  & 7.28(2)   & 1.94(3) \\
$\mathrm{\Phi_{pp-1}}$  & 5.9(-1)  & 6.46(1)  & 2.21(1)   & 3.43(2) & 3.66(2) &1.44(1)  & 2.55(2)   & 1.38(2) \\
$\mathrm{\Phi_{^{7}Be}}$& 5.8(-3)  & 3.38(-3) & 7.35(-5)  & 1.26(-1)& 9.90(1)&1.44(-1) & 6.91(1)   & 4.78(1) \\
$\mathrm{\Phi_{^{8}B}}$ & 1.4(-2)  & 2.87(-4) & 1.74(-8)  & 4.76(-3)& 1.07(1) &1.68(-1) & 4.29      & 1.16(-1)\\
$\mathrm{\Phi_{^{13}N}}$& 6.0(-2)  & 8.03     & 2.28(-1)  & 3.42(2) & 5.74(-1)&1.83(1). & 2.00(2)   & 8.75(2) \\
$\mathrm{\Phi_{^{15}O}}$& 5.0(-2)  & 7.58     & 1.69(-2)  & 3.42(2) & 3.63(2)&6.12(-1) & 1.97(2)   & 8.74(2) \\
$\mathrm{\Phi_{^{17}F}}$& 6.0(-4)  & 8.17(-2) & 1.61(-5)  & 2.66    & 1.46 &1.00(-2) & 1.09      & 7.74(-2)\\
$\mathrm{\Phi_{pep}}$   & 1.42(-1) & 1.42(-1) & 1.42(-1)  & 4.54(-2)& 7.835(2) &2.00(-2) & 3.93(-1)  & 9.68(-2)\\
\hline
\end{tabular}
\protect\caption[]{\footnotesize{}
{Energy spectrum for the stars considered in this work. On each graph the dashed lines represents the contribution from the different processes in the solar neutrino spectrum, using the standard solar model by \cite{BP2004}. Table at the end displays the neutrino flux, contributions of the 
neutrino flux, were calculated as if each star was located at a distance of 1 A.U.}}
\end{figure*}

\subsubsection{Sirius A}

Sirius A is the brightest member in the $\alpha$-CMa system, whose other component is Sirius B, a well known white dwarf. The calibrations for its observational stellar parameters, \cite{Sir1,Sir3}, are shown on the upper part of the fourth column in table II. One of the most notorious characteristics of Sirius is the metallicity of its surface, equivalent to $\mathrm{Z=0.063}$, more metallic than any star within 10 pc from the Sun \cite{Sir3}. According to \cite{Sir2} this enhanced 
metallicity is not representative for the interior, as it could be produced by radiation pressure lifting up metals from its inner zones, suggesting that $\mathrm{Z=0.012}$ could be a suitable choice for the metallicity of Sirius. Hence this value was adopted
for the simulation.
The model for Sirius is shown in table II. The predicted temperature and density of the stellar core are: 
$\mathrm{T_{c}=22.25}$ and $\mathrm{\rho_{c}=0.68}$, where the low density of the core could be a consequence of convection being the main source for heat transfer. Neutrino luminosity is around 61 times the 
solar neutrino luminosity and 70\% comes from the CNO-cycle.  
The right middle panel on fig. 2 shows the neutrino spectrum for Sirius A, $\mathrm{\Phi_{T}=120 \times10^{10}\cdot cm^{-2}\cdot s^{-1}}$, 20 times the flux of neutrinos from the Sun. The spectral distribution is very similar to that of Procyon A (the same dominant reactions producing almost 90$\%$ of the total 
flux). The largest differences are the scale of the neutrino flux from the $\mathrm{^{7}Be}$ and $\mathrm{^{8}B}$ reactions, as they are three orders of magnitude larger than their solar analogs. An enhanced flux from the $\mathrm{^{8}B}$ could be a common characteristic on stars in which core hydrogen burning is dominated by the CNO-cycle (as can be seen also on the spectra for Altair and Vega). The increased core 
temperature results in an increase of $\mathrm{^{8}B}$ neutrinos.

\subsubsection{Altair}

Altair is an A7V star in the Eagle constellation, at 5.13 pc from the solar system, with an approximate age of $1.26\times10^{9}$ yrs \cite{altair2}. Along with Vega, Altair is a fast rotator, with an estimated rotational speed of $\mathrm{240}$ km$\mathrm{\cdot s^{-1}}$ \cite{altair}. This makes its equator to be around $\mathrm{2 R_{\odot}}$ wide while its poles are about $\mathrm{1.63R_{\odot}}$. Altair's rotation also affects the determination of its effective temperature, the best range is set between 6900 to 8500 K \cite{altair3}, the model
described here is using 7860 K. 
The stellar model is shown on table II. As in previous models, the stellar track was stopped as soon as it matched the observational data reported in \cite{altair3}. Although the lack of spherical symmetry, one of the assumptions made by the Eggleton code, implies that the predictions by the stellar model can only be taken as approximations, its effects on luminosity should be minimum as, according to \cite{altair3}, this parameter its more sensitive to other physical parameters. Altair's neutrino luminosity is around 10 times the solar value, and almost 73\% is produced by the CNO-cycle.
The total neutrino flux from Altair ($\mathrm{\Phi_{T}=33.63\times10^{10} \cdot cm^{-2}\cdot s^{-1}}$) is shown on the right panel at the bottom on figure II. The largest flux corresponds to the $\mathrm{^{13}N}$ reaction (54\%), followed by the pp-I reaction (43\%). The other reactions from the CNO-cycle produce most of the remaining neutrinos, one order of magnitude larger than in the Sun. Like Sirius, the flux of 
$\mathrm{^{8}B}$ neutrinos is larger than that of the Sun. The peak emission of the $\mathrm{^{13}N}$ and $\mathrm{^{15}O}$, unlike to what is displayed for Altair's spectrum, are almost equal but lower to those shown for Sirius.

\subsubsection{Vega}

Vega (spectral class A0V) is the brightest star in the Lyra constellation, the most luminous among the stars considered in this work, with around 40 times the solar bolometric luminosity. Recent studies suggest that Vega could be at the middle of its main-sequence (with a current age of $4.55\times10^{8}$ years). As Altair, 
Vega's shape is oblate due to its high rotational velocity ($\mathrm{236 km 
s^{-1}}$) reaching $\mathrm{2.81 R_{\odot}}$ on its equator and 
$\mathrm{2.36 R_{\odot}}$ on its pole and, as a consequence, there are variations in the effective temperature ranging from 6900 to 8500 K. Hence, as for Altair's model, the results can be taken only as approximate.\\
The last column in table II shows the stellar model for Vega, with a bolometric luminosity $\mathrm{L=39.87 L_{\odot}}$, effective temperature of 8500K and surface radius of $\mathrm{2.96 R_{\odot}}$. Vega's model has the hottest temperature among the stars in this work, $\mathrm{T_{c}=28.44}$. Neutrino luminosity is around 83 times that of the Sun and almost 90\% is due to the reactions of the CNO-cycle.
The neutrino spectrum of Vega is shown in the last panel on figure 2. The flux of neutrinos ($\mathrm{\Phi_{T}=103\times10^{10}\cdot cm^{-2}\cdot s^{-1}}$) is dominated by the CNO reactions, being between two and three orders of magnitude larger than their solar analogs. Similarly, the flux of neutrinos from the $\mathrm{^{8}B}$ reaction is about two hundred times larger. The neutrino spectrum of Vega is the only one in which the peak of the flux of neutrinos for the pp-I reaction is not the highest, it is below those for the $\mathrm{^{13}N}$ and $\mathrm{^{15}O}$ reactions.

\subsubsection{Fomalhaut}

Fomalhaut is the brightest star in the Piscis Austrinus constellation, part of a triple stellar system whose other members are a K star and an M dwarf, both too weak to contribute in a significant way to the neutrino output from the system. 
Fomalhaut is still at its early main-sequence, with an approximate age of 
$4\times10^{8}$ yrs, and its mass has been estimated to 
$\mathrm{M=1.92M_{\odot}}$ \cite{Fomalhaut1}. There is a general agreement about the metallicity of Fomalhaut being sub-solar but the estimations have gone from 
$\mathrm{[Fe/H]}=-0.03$ \cite{Fomalhaut2} to $\mathrm{[Fe/H]=-0.34}$ \cite{Fomalhaut3}.
The first column in table II shows Fomalhaut's stellar model, using the most recent estimation for its metallicity \cite{Fomalhaut3}. The density and temperature at the stellar core are very similar to those of Sirius ($\mathrm{\rho_{c}=0.72}$ and $\mathrm{T_{c}=21.53}$), slightly cooler but hot enough to favor the reactions of the CNO-cycle over the pp-chain. The stellar model predicts a neutrino luminosity of $\mathrm{35.40 L_{\odot}}$, from which 70\% of the neutrinos are produced by the CNO-cycle.
Fomalhaut's spectrum is not shown in fig. 2 due to its similarity to those for Vega and Sirius, the magnitude of the flux from each reaction being the only difference but having the same proportions as with their analogs for the other A stars. According to the stellar model, the total neutrino flux from Fomalhaut should be around $\mathrm{\Phi_{T}=72.7\times10^{10} \cdot cm^{-2}\cdot s^{-1}}$, with 55\% of the neutrinos being produced by the reactions in the CNO cycle, while the pp-I, $\mathrm{^{7}Be}$ $\mathrm{^{8}B}$ and pep-reactions contribute with 35\%, 9.5\%, 0.50\% and 0.0054\%, respectively.

\section{Detection opportunity}

\begin{table}
\scriptsize
\begin{tabular}{|l c c c|}
	\hline
source             & rate  & $\mathrm{\Phi_{1A.U}}$  & $\mathrm{\Phi_{true}}$   \\
 & $[\mathrm{10^{38}\nu\cdot s^{-1}}]$ & $[10^{10}s^{-1}\cdot cm^{-2}]$ & $[\nu\cdot s^{-1}\cdot cm^{-2}]$\\
\hline 
Sirius A           & 23.715           &  120.64 & 4.05 \\
Procyon A          & 21.941           &  102.97 & 1.97 \\
$\alpha$-Centauri A\&B  &  1.122           &   10.28 & 1.43 \\
Vega               & 44.154           &  193.64 & 0.76 \\
Fomalhaut          & 13.730           &   72.76 & 0.30 \\
Altair             &  5.242           &   33.63 & 0.29 \\
\hline
\end{tabular}
\protect\caption[]{\footnotesize{Stars considered in this work and their neutrino production rate (first column), what would be their flux as if they were located at 1 AU (second column) and their predicted real fluxes at Earth (third column).}}
\end{table}

The neutrino flux at Earth can be determined using the actual distance to 
each star (see Table III). The integrated fluxes expected at Earth are only of the order of 1-10 $\mathrm{\nu \cdot s^{-1}\cdot 
cm^{-2}}$, the major contribution comes from the neutrinos produced by the CNO-cycle of the considered stars.
This number is in the same region as limits Super-Kamiokande has already obtained for the DSNB, but
CNO-neutrinos are below their detection threshold. New projects are proposed or under construction
like Hyper-Kamiokande, a 1 Mton water Cerenkov detector \cite{sno}, several 
scintillator detectors like JUNO  (20 kton) \cite{juno}, LENA (50 kton) 
\cite{wur11}, RENO-50 (18 kton) \cite{reno-50} and the
Jinping experiment \cite{bea16} planing a 10 times larger fiducial volume (about 4 kt) than Borexino deep underground. 
Due to their threshold water Cerenkov detectors are only sensitive to $^{8}$B neutrinos but give 
directional information, on the other hand liquid scintillators have a very low 
energy threshold down into the pp-region but not directional information.
Hence most promising way, would be large scale water based scintillators \cite{li}, which
is considered for the Jinping experiment.
The fast Cerenkov light from the water component is guaranteeing some directional information 
and the scintillator allows for low thresholds. The Theia-project is also
planning to work in this direction using several kilotons of water based 
liquid scintillator underground \cite{theia}.
\\   
Given the very low flux an observation might be very difficult by neutrino-electron scattering. 
The major background will be solar neutrinos. Therefore  
directional information is mandatory to discriminate events from background. 
Taking the directional resolution for solar neutrinos above 5 MeV, there will be no overlap of the 
discussed candidate stars with the Sun. Run periods were the Sun might be 
close to one of the stars could be rejected. Also a relative measurement of event rates within
nearby regions in the surrounding areas of the candidate star could be used to
determine background and might allow to search for a small signal.  
A potential detection of thermal neutrinos is very unlikely due to their low energy and flux.

\section{Summary and conclusions}
The aim of this paper is to explore the potential of detecting neutrino from stars
in the solar neighbourhood, considered are stars within 10 pc from the Earth. 
The major contributors were simulated by using the Eggleton stellar evolution code. 
The estimated neutrino fluxes are just about in the region of detection
of current experiments. Next generation experiments should be
able to detect these neutrinos, however directional information will be vital 
to remove the dominant solar neutrino background.

\section{Acknowledgement} 
The authors would like to thank Dr Xuefei Chen and Christopher A. Tout for their OPAL tables, implemented in our version of the Eggleton code.

\end{document}